\documentclass[12pt]{article}

\input{tcilatex}

\begin{document}

\begin{center}
\textbf{PROPAGATION OF EXTREMELY SHORT PULSES IN NON-RESONANT MEDIA : THE
TOTAL MAXWELL-DUFFING MODEL }

\bigskip

\vspace{0.5in}Andrei I. Maimistov$^{a}$ \footnote{%
electronic address: maimistov@pico.mephi.ru} and Jean-Guy Caputo$^{b,c}$ 
\label{a}\footnote{%
electronic address: caputo@insa-rouen.fr}

\vspace{0.2cm} $^{a}$ Department of Solid State Physics, Moscow Engineering
Physics Institute, Kashirskoe sh. 31, Moscow, 115409 Russia

$^{b}$ Laboratoire de Math\'{e}matiques, INSA de Rouen, B.P. 8, 76131
Mont-Saint-Aignan cedex, France \vspace{0.2cm}

$^{c}$ Laboratoire de Physique th\'eorique et modelisation, Universit\'e de
Cergy-Pontoise and C.N.R.S.
\end{center}

\bigskip

\begin{center}
\textbf{ABSTRACT}
\end{center}

\bigskip

Propagation of extremely short pulses of electromagnetic field
(electromagnetic spikes) is considered in the framework of the total
Maxwell-Duffing model where anharmonic oscillators with cubic nonlinearities
(Duffing model) represent the material medium and wave propagation is
governed by the 1-d bidirectional Maxwell equations. This system of
equations has a one parameter family of exact analytical solutions
representing an electromagnetic spike propagating on a zero or a nonzero
background. We find that the total Maxwell-Duffing equations can be written
as a system in bilinear form and that the one-soliton solution of this
system coincides with the steady state solution obtained previously.



\section{\protect\LARGE Introduction}

Ultra short nonlinear pulses of an electromagnetic field, which contain as
few as one half optical cycle, have recently attracted a great deal of
attention \cite{R1}-\cite{R5}. The description of the evolution of the
electromagnetic field was based on Maxwell equations or the subsequent wave
equation. To describe a medium where electromagnetic waves propagate, one
frequently uses an ensemble of oscillators or resonant atoms. If the
oscillators are linear we obtain the important Lorentz model, which has been
very useful to describe the propagation of an electromagnetic wave in a
linear medium. The simplest generalization of the Lorentz model is obtained
by adding an anharmonic term to the equation of the oscillator. This leads
to the Duffing model in the case of a cubic nonlinearity \cite{R6,R7,R8}. In
the framework of the Duffing model it was shown that solitary pulses of a
unidirectional electromagnetic field (''electromagnetic bubbles'') having an
amplitude comparable to the atomic field and a duration down to $\sim
10^{-16}$ s may be expected. These bubbles propagate without dispersion as
stable solitary waves. Very short pulses of this kind will be referred to as 
\emph{extremely short pulses} (ESP).

In all these works, the oscillator represents the response of the
high-frequency electronic degree of freedom to the electromagnetic field.
The steady state ESP, which have been found heretofore, can be represented
by a moving wave packet with a zero dominant frequency. Some authors call
such pulses video pulses \cite{Sazo, Shvarts} to emphasize the difference
from the second kind of ESP with a non-zero dominant frequency. These
(optical) ESP are considered in Ref. \cite{R5}. Hereafter we will consider
ESP's of the first kind. Since the ESP spectrum is concentrated at low
frequencies, ion oscillations may also give a considerable contribution to
the full polarization of the medium. Nevertheless, we will neglect the ion
component in the material response (the propagation of femtosecond pulses in
a medium with a nonlinearity determined by both electronic and ionic
(Raman-scattering) degrees of freedom was considered in Ref. \cite{R9}).

As the Duffing model is the simplest and most fruitful generalization of the
Lorentz model, its investigation is very attractive. An objective of the
present work is to study the propagation of linearly polarized ESPs in a
nonlinear dispersive medium modeled by an anharmonic oscillator
characterized by the cubic nonlinearities. The evolution of the electric
field of ESP will be considered on the base of the Maxwell equations without
any unidirectional reduction.

The paper is structured as follows. The model is derived in section 2
together with dynamical invariants or integrals of motion. Two families of
moving ESP analytical solutions are given in section 3. In section 4 the
bilinear form of these equations will be presented and we conclude in
section 5.

\section{\protect\LARGE Constitutive model}

To consider the propagation of an extremely short electromagnetic pulse
propagation in a nonlinear dispersive medium we need to use the total
Maxwell wave equation and some model for the medium. Let us consider the
standard Lagrangian density for an electromagnetic field taking into account
the interaction with an ensemble of nonlinear oscillators: 
\begin{equation}
\mathcal{L}=\frac{1}{2c_{0}^{2}}A_{,t}^{2}-\frac{1}{2}A_{,z}^{2}+4\pi
\sum\limits_{a}\left\{ \frac{1}{2}mX_{a,t}^{2}-\frac{1}{2}mX_{a}^{2}-\frac{1%
}{4}\kappa _{3a}X_{a}^{4}-\frac{e_{\ast }}{c_{0}}AX_{a,t}\right\}
\label{eq1}
\end{equation}

Here we consider a plane electromagnetic wave, propagating along the $z$%
-axis and represented by the vector potential $A$. We use an anharmonic
oscillators model (Duffing model) to reproduce the electronic response of an
atom located at the spatial point indicated by the symbol $a$. The electrons
are considered as particles in a potential well characterized by the
displacements from their equilibrium position $X_{a}$. They oscillate with
their own frequencies $\omega _{a}$ and are influenced by the
electromagnetic field. In expression (\ref{eq1}) $e_{\ast }$ is the electric
charge of the electron and $\kappa _{3a}$\ are anharmonicity coefficients.
Hereafter, we will use $m$ as a symbol for this effective mass, which
accounts for the local Lorentz field effect. The partial derivatives we will
denote as $\partial f/\partial t=f_{,t},\partial f/\partial z=f_{,z}$ and so
on.

The application of the variational procedure to the action related with the
Lagrangian density (\ref{eq1}) yields equations 
\[
A_{,zz}-c^{-2}A_{,tt}=(4\pi e_{\ast }/c_{0})\sum_{a}X_{a,t}, 
\]
\[
mX_{a,tt}+m\omega _{a}^{2}X_{a}+\kappa _{3a}X_{a}^{3}=(e_{\ast
}/c_{0})A_{,t} 
\]

If one introduces the strength of the electric field\ $E=c_{0}^{-1}A_{,t}$,
then the constituent equations of the model under consideration can be
written as

\begin{equation}
E_{,zz}-c^{-2}E_{,tt}=(4\pi /c_{0}^{2})P_{,tt}  \label{eq2_1}
\end{equation}
\begin{equation}
X_{a,tt}+\omega _{a}^{2}X_{a}+(\kappa _{3a}/m)X_{a}^{3}=(e_{\ast }/m)E
\label{eq2_2}
\end{equation}
The polarization $P$ of the nonlinear medium is $P=\sum_{a}e_{\ast }X_{a}$ .

Let us consider the case of a homogeneous broadening medium where all atoms
have the same parameters, in particular $\omega _{a}=\omega _{0}$. Then we
can write the polarization as $P=n_{A}e_{\ast }X$, where $n_{A}$\ is the
density of the oscillators (atoms), and the index of the atom can be omitted.

We rescale the variables and fields as $\zeta =z\omega _{0}/c_{0}$, $\tau
=\omega _{0}t$\ and $e=E/E_{0}$, $q=X/X_{0}$, where 
\[
E_{0}=m\omega _{0}^{2}X_{0}/e_{\ast }=m\omega _{0}^{3}e_{\ast }^{-1}(2\mu
/|\kappa _{3}|)^{1/2},\quad X_{0}=(2\mu m\omega _{0}^{2}/|\kappa
_{3}|)^{1/2},
\]
and we will also use the following parameters $\gamma =\omega _{p}/\omega
_{0}$, $2\mu =\kappa _{3}X_{0}^{2}/m\omega _{0}^{2}$, where $\omega
_{p}=(4\pi n_{A}e_{\ast }^{2}/m)^{1/2}$ is the plasma frequency. In terms of
the rescaled variables equations (\ref{eq2_1}) and (\ref{eq2_2}) take the
form 
\begin{equation}
e_{,\zeta \zeta }-e_{,\tau \tau }=\gamma q_{,\tau \tau }~~,~~q_{,\tau \tau
}+q+2\mu q^{3}=\gamma e.  \label{eq3}
\end{equation}
These equations will be named the \emph{total Maxwell-Duffing equations} (or
TMD-equations). To this system should be added the initial conditions: 
\[
e(\zeta =0,\tau )=e_{0}(\tau ),\quad e_{,\tau }(\zeta =0,\tau
)=e_{0}^{\prime }(\tau ),
\]
and the boundary conditions 
\[
q(\zeta ,\tau )=q_{,\tau }(\zeta ,\tau )=0,\quad \mathrm{at}~~\tau
\rightarrow \pm \infty .
\]
related to the evolution of an initial electromagnetic solitary wave in a
nonlinear dispersive medium. 

Note that the system of equations (\ref{eq3}) can be rewritten in
alternative form by introducing the new auxiliary field variable $b$ by the
relation $\partial e/\partial \zeta =\partial b/\partial \tau $. In this
case the TMD-equations take the following form: 
\begin{equation}
e_{,\zeta }=b_{,\tau }~~,~~ b_{,\zeta }=e_{,\tau }+\gamma p~~,~~q_{,\tau
}=p~~,~~p_{,\tau }+q+2\mu q^{3}=\gamma e.  \label{eq4}
\end{equation}

It should be noted that the system of equations (\ref{eq3}) can be derived
as the Euler-Lagrange equations from the action functional 
\[
S=\int \mathcal{L}[q,a]d\tau d\zeta , 
\]
where now the new Lagrangian density is 
\begin{equation}
\mathcal{L}=\frac{1}{2}\left( \frac{\partial a}{\partial \zeta }\right) ^{2}-%
\frac{1}{2}\left( \frac{\partial a}{\partial \tau }\right) ^{2}-\frac{1}{2}%
\left( \frac{\partial q}{\partial \tau }\right) ^{2}+\frac{1}{2}q^{2}+\frac{%
\mu }{2}q^{4}-\gamma q\frac{\partial a}{\partial \tau }.  \label{eq5}
\end{equation}
Applying the variational procedure to the action $S$ yields equations 
\begin{equation}
\frac{\partial ^{2}a}{\partial \zeta ^{2}}-\frac{\partial ^{2}a}{\partial
\tau ^{2}}=\gamma \frac{\partial q}{\partial \tau }~~,~~\frac{\partial ^{2}q%
}{\partial \tau ^{2}}+q+2\mu q^{3}=\gamma \frac{\partial a}{\partial \tau }
\label{eq6}
\end{equation}
Identifying $a$\ as a potential for the field $e$, so that $e=a_{,\tau }$,
makes these equations identical to Eqs. (\ref{eq3}).

From the Lagrangian density (\ref{eq5}) we can get the density of moments of
the fields $a$ and $q$: 
\begin{equation}
\pi _{a}(\zeta ,\tau )=\frac{\partial \mathcal{L}}{\partial a_{,\zeta }}%
=a_{,\zeta }(\zeta ,\tau )=b(\zeta ,\tau )~~,~~\pi _{q}(\zeta ,\tau )=\frac{%
\partial \mathcal{L}}{\partial q_{,\zeta }}=0.  \label{eq7}
\end{equation}
The second expression in (\ref{eq7}) indicates that we have a degenerate
Lagrangian, which leads to a constrained Hamiltonian system ($\pi _{q}(\zeta
,\tau )=0$ \ is a primary constraint) \cite{R10,R11}.

Now one can get a canonical Hamiltonian density using the Legendre transform 
\begin{equation}
\mathcal{H}=a_{,\zeta }\frac{\partial \mathcal{L}}{\partial a_{,\zeta }}-%
\mathcal{L}=\frac{1}{2}\left( a_{,\zeta }^{2}+a_{,\tau }^{2}+q_{,\tau
}^{2}-q^{2}-\mu q^{4}+2\gamma qa_{,\tau }\right) .  \label{eq8}
\end{equation}
The integration of this expression with respect to $\tau $ leads to the
canonical Hamiltonian which is an integral of motion 
\begin{equation}
H=\frac{1}{2}\int\limits_{-\infty }^{\infty }\left( e^{2}+b^{2}+q_{,\tau
}^{2}-q^{2}-\mu q^{4}+2\gamma qe\right) d\tau .  \label{eq9}
\end{equation}

It is worth noting that there are two additional integrals of motion, which
follow from the TMD-equations (\ref{eq4}): 
\begin{equation}
I_{1}=\int\limits_{-\infty }^{\infty }e(\zeta ,\tau )d\tau
=\int\limits_{-\infty }^{\infty }a_{,\tau }(\zeta ,\tau )d\tau =a(\zeta
,\tau =\infty )-a(\zeta ,\tau =-\infty ),  \label{eq10}
\end{equation}
\noindent and 
\begin{equation}
I_{2}=\int\limits_{-\infty }^{\infty }b(\zeta ,\tau )d\tau
=\int\limits_{-\infty }^{\infty }\pi _{a}(\zeta ,\tau )d\tau  \label{eq11}
\end{equation}
The magnitude of the first integral is defined by the boundary conditions
only so that it can be interpreted as a topological charge in the
Maxwell-Duffing model. The second integral is the canonical moment in this
model.

\section{\protect\LARGE Steady state solutions}

Let us look for solutions of the TMD-equations as traveling waves with a
non-varying profile. First consider solitary waves on a zero background,
i.e. assume the following boundary conditions at $\tau \rightarrow \pm
\infty $: 
\[
e(\zeta ,\tau )=e_{,\tau }(\zeta ,\tau )=0~~,~~q(\zeta ,\tau )=q_{,\tau
}(\zeta ,\tau )=0,
\]
Substituting $e(\zeta ,\tau )=e(\tau -\zeta /v)$,~~ $q(\zeta ,\tau )=q(\tau
-\zeta /v)$ into equation (\ref{eq3}) and taking into account the boundary
conditions one finds that the first equation of (\ref{eq3}) results in $%
e=\alpha q/\gamma $, where $\alpha =\gamma ^{2}v^{2}/(1-v^{2})$. The second
equation of (\ref{eq3}) can be transformed into the following ordinary
differential equation in the variable $T=\tau -\zeta /v$ 
\[
\frac{d^{2}q}{dT^{2}}-(\alpha -1)q+2\mu q^{3}=0.
\]
A non-singular solution of this equation exists only if $\alpha >1$\ and $%
\mu >0$. In this case integrating results in the following expression \cite
{R2,R7,R8} 
\begin{equation}
q_{st}(\zeta ,\tau )=\pm \sqrt{(\alpha -1)/\mu }\sec \mathrm{h}\left[ \sqrt{%
(\alpha -1)}\left( \tau -\zeta /v-\tau _{0}\right) \right] .  \label{eq12}
\end{equation}
Here the integration constants $\tau _{0}$ and $\alpha $\ are the parameters
of this steady state solitary wave. The strength of the electric field of
the ESP is given by the following formula 
\begin{equation}
e_{st}(\zeta ,\tau )=\pm \alpha \gamma ^{-1}\sqrt{(\alpha -1)/\mu }\sec 
\mathrm{h}\left[ \sqrt{(\alpha -1)}\left( \tau -\zeta /v-\tau _{0}\right) %
\right]   \label{eq13}
\end{equation}
This solution describes the propagation of an electromagnetic spike with
positive ($+$ sign ) or negative ($-$ sign ) polarity. The condition for
existence of this solution leads to the limitation of the velocity: $%
1<v<(1+\gamma ^{2})^{-1/2}$.

Now let us consider solitary waves on a non-zero background, i.e. we assume
the following boundary conditions at $\tau \rightarrow \pm \infty $:: 
\begin{equation}
e(\zeta ,\tau )=e_{0},\quad q(\zeta ,\tau )=q_{0}~~,~~e_{,\tau }(\zeta ,\tau
)=0~~,~~q_{,\tau }(\zeta ,\tau )=0,  \label{eq14a}
\end{equation}
\begin{equation}
\gamma e_{0}=q_{0}+2\mu q_{0}^{3}  \label{eq14b}
\end{equation}
Introduce the following fields $u(\zeta ,\tau )=e(\zeta ,\tau )-e_{0}$\ \
and $f(\zeta ,\tau )=q(\zeta ,\tau )-q_{0}$. Instead of equations (\ref{eq3}%
) we obtain the new system of equations 
\begin{equation}
\frac{\partial ^{2}u}{\partial \zeta ^{2}}-\frac{\partial ^{2}u}{\partial
\tau ^{2}}=\gamma \frac{\partial f}{\partial \tau }~~,~~\frac{\partial ^{2}f%
}{\partial \tau ^{2}}+(1+6\mu q_{0}^{2})f+6\mu q_{0}f^{2}+2\mu f^{3}=\gamma 
\frac{\partial u}{\partial \tau }.  \label{eq15}
\end{equation}
Under the assumption that $u(\zeta ,\tau )=u(\tau -\zeta /v)$, $f(\zeta
,\tau )=f(\tau -\zeta /v)$ \ the first equation leads to the relation $%
e=\alpha q/\gamma $, and the second equation takes the form 
\begin{equation}
\frac{d^{2}f}{dT^{2}}=(\alpha -1-6\mu q_{0}^{2})f-6\mu q_{0}f^{2}-2\mu f^{3}.
\label{eq16}
\end{equation}
From (\ref{eq16}) and the boundary conditions (\ref{eq14a})\ it follows that 
\begin{equation}
\left( \frac{df}{dT}\right) ^{2}=(\alpha -1-6\mu q_{0}^{2})f^{2}-4\mu
q_{0}f^{3}-\mu f^{4}.  \label{eq17}
\end{equation}
A non-singular solution of this equation exists when $\alpha _{1} \equiv
\alpha -6\mu q_{0}^{2}>1$ and $\mu >0$. This condition leads to the
limitation of the velocity of the ESP 
\[
1<v^{2}<v_{c}^{2}=\frac{1+6\mu q_{0}^{2}}{1+\gamma ^{2}+6\mu q_{0}^{2}}. 
\]
The solution of equation (\ref{eq17}) can be written as 
\begin{equation}
f^{(\pm )}(\zeta ,\tau )=\frac{\pm (\alpha _{1}-1)}{\sqrt{(2\mu
q_{0})^{2}+\mu (\alpha _{1}-1)^{2}}\cosh [(\alpha _{1}-1)^{1/2}(\tau -\zeta
/v-\tau _{0})]\pm 2\mu q_{0}}  \label{eq18}
\end{equation}
The steady state pulse of electromagnetic wave propagating on a constant
electric background is then given by 
\begin{equation}
e^{(\pm )}(\zeta ,\tau )=e_{0}\pm \frac{\alpha (\alpha _{1}-1)}{\sqrt{(2\mu
q_{0})^{2}+\mu (\alpha _{1}-1)^{2}}\cosh [(\alpha _{1}-1)^{1/2}(\tau -\zeta
/v-\tau _{0})]\pm 2\mu q_{0}}  \label{eq19}
\end{equation}
Unlike the case of ESP propagation without background, here ESPs of
different polarities have different amplitudes.

\section{\protect\LARGE Bilinear form of the total Maxwell-Duffing equations}

If the following substitutions 
\begin{equation}
e=\frac{g}{h}~~,~~b=\frac{c}{h}~~,~~q=\frac{f}{h}  \label{eq20}
\end{equation}
are used, then equations (\ref{eq4}) can be rewritten as 
\[
\frac{1}{h^{2}}D_{\zeta }(g\cdot h)-\frac{1}{h^{2}}D_{\tau }(c\cdot h)=0, 
\]
\begin{equation}
\frac{1}{h^{2}}D_{\zeta }(c\cdot h)-\frac{1}{h^{2}}D_{\tau }(g\cdot
h)=\gamma \frac{1}{h^{2}}D_{\tau }(f\cdot h),  \label{eq21}
\end{equation}
\[
\frac{1}{h^{2}}D_{\tau }^{2}(f\cdot h)-\frac{f}{h^{3}}D_{\tau }^{2}(h\cdot
h)+\frac{f}{h}+2\mu \frac{f^{3}}{h^{3}}=\gamma \frac{g}{h}, 
\]
where the Hirota $D$-operators $D_{\zeta }(a\cdot b)=a_{,\zeta }b-ab_{,\zeta
}$, $D_{\tau }(a\cdot b)=a_{,\tau }b-ab_{,\tau }$ have been introduced \cite
{R12}. To derive the last equation we follow the rule 
\[
\frac{\partial ^{2}}{\partial \tau ^{2}}\left( \frac{f}{h}\right) =\frac{1}{%
h^{2}}D_{\tau }^{2}(f\cdot h)-\frac{f}{h^{3}}D_{\tau }^{2}(h\cdot h). 
\]
Multiplying the first equation by $h^{2}$, the second equation one by $h^{3}$%
, the third one by $h^{2}$ and collecting the terms of order $h^{-1}$, we
can write the resulting equations as a system of bilinear ones 
\begin{eqnarray}
D_{\zeta }(g\cdot h) &=&D_{\tau }(c\cdot h),  \nonumber \\
D_{\zeta }(c\cdot h) &=&D_{\tau }(g\cdot h)+\gamma D_{\tau }(f\cdot h),
\label{eq22} \\
D_{\tau }^{2}(f\cdot h) &=&(\gamma g-f)h,  \nonumber \\
D_{\tau }^{2}(h\cdot h) &=&2\mu f^{2}  \nonumber
\end{eqnarray}

We will use the usual method \cite{R12, R13}\ to solve equations (\ref{eq22}%
) by writing 
\begin{equation}
g=g_{1}e^{\theta },\quad c=c_{1}e^{\theta },\quad f=f_{1}e^{\theta },\quad
h=1+h_{1}e^{\theta }+h_{2}e^{2\theta },  \label{eq23}
\end{equation}
where $\theta =k\zeta -\Omega \tau $. Substituting this into (\ref{eq22})
results in the equations: 
\begin{eqnarray*}
kg_{1}e^{\theta }-kh_{2}g_{1}e^{3\theta }+\Omega c_{1}e^{\theta }-\Omega
h_{2}c_{1}e^{3\theta } &=&0, \\
kc_{1}e^{\theta }-kh_{2}c_{1}e^{3\theta }+\Omega g_{1}e^{\theta }-\Omega
h_{2}g_{1}e^{3\theta }+\Omega \gamma f_{1}e^{\theta }-\Omega \gamma
h_{2}f_{1}e^{3\theta } &=&0, \\
\Omega ^{2}f_{1}e^{\theta }+\Omega ^{2}h_{2}f_{1}e^{3\theta }-(\gamma
g_{1}-f_{1})(e^{\theta }+h_{1}e^{2\theta }+h_{2}e^{3\theta }) &=&0, \\
\Omega ^{2}h_{1}e^{\theta }+4\Omega ^{2}h_{2}e^{2\theta }+\Omega
^{2}h_{2}h_{1}e^{3\theta }-\mu f_{1}^{2}e^{2\theta } &=&0.
\end{eqnarray*}
Equating the coefficients of the different powers of $e^{\theta }$\ to zero,
one obtains the system of equations that define $%
c_{1},f_{1},g_{1},h_{1,}h_{2}$ : 
\begin{eqnarray*}
kg_{1}+\Omega c_{1} &=&0~~,~~kc_{1}+\Omega (g_{1}+\gamma f_{1})=0,\quad \\
\Omega ^{2}f_{1}-(\gamma g_{1}-f_{1}) &=&0~~,~~4\Omega ^{2}h_{2}=\mu
f_{1}^{2}~~,~~h_{1}=0.
\end{eqnarray*}
From these relations one can get 
\begin{eqnarray*}
h_{1} &=&0~~,~~h_{2}=\mu f_{1}^{2}(2\Omega )^{-2}, \\
c_{1} &=&-\gamma ^{-1}(k/\Omega )(1+\Omega ^{2})f_{1}~~,~~g_{1}=\gamma
^{-1}(1+\Omega ^{2})f_{1},
\end{eqnarray*}
and the ''dispersion relation'' 
\begin{equation}
k^{2}=\frac{\Omega ^{2}(1+\gamma ^{2}+\Omega ^{2})}{(1+\Omega ^{2})}.
\label{eq24}
\end{equation}
It should be pointed out that in the low frequency limit this formula yields 
$k^{2}=\Omega ^{2}(1+\gamma ^{2})$, while in the high-frequency limit it
yields $k^{2}=\Omega ^{2}$.

Thus, we found a one-soliton solution of the bilinear equations (\ref{eq22})
which can be written as 
\begin{eqnarray*}
g &=&\gamma ^{-1}(1+\Omega ^{2})f_{1}e^{\theta },\quad \\
c &=&-\gamma ^{-1}(k/\Omega )(1+\Omega ^{2})f_{1}e^{\theta }, \\
f &=&f_{1}e^{\theta }~~,\quad h=1+\mu f_{1}^{2}(2\Omega )^{-2}e^{2\theta }.
\end{eqnarray*}
These relations yield a solution of the TMD-equations (\ref{eq5}) which is
consistent with the steady state solution obtained earlier: 
\begin{eqnarray}
e_{sol}(\zeta ,\tau ) &=&\frac{\gamma ^{-1}(1+\Omega ^{2})f_{1}e^{\theta }}{%
1+\mu (f/2\Omega )^{2}e^{2\theta }}, \\
b_{sol}(\zeta ,\tau ) &=&-\frac{\gamma ^{-1}(k/\Omega )(1+\Omega
^{2})f_{1}e^{\theta }}{1+\mu (f/2\Omega )^{2}e^{2\theta }},  \label{eq25} \\
q_{sol}(\zeta ,\tau ) &=&\frac{f_{1}e^{\theta }}{1+\mu (f/2\Omega
)^{2}e^{2\theta }}.  \nonumber
\end{eqnarray}
If we introduce a new constant of integration $\exp \theta _{0}=\pm \mu
^{1/2}f_{1}/2\Omega $, then the one-soliton expression for the normalized
electric field of ESP can be written as 
\begin{equation}
e_{sol}(\zeta ,\tau )=\pm \left( \frac{1+\Omega ^{2}}{\mu ^{1/2}\gamma }%
\right) \frac{\Omega }{\cosh (\theta +\theta _{0})}.  \label{eq26}
\end{equation}
The velocity of this soliton is defined as $v=\Omega /k$. Taking into
account the dispersion relation (\ref{eq24}) one can obtain 
\[
v^{2}=\frac{1+\Omega ^{2}}{1+\gamma ^{2}+\Omega ^{2}}. 
\]
The magnitude of this velocity lies in the interval\ $(1,(1+\gamma
^{2})^{-1/2}).$\ \ This is the same result as for a steady state pulse
obtained above. Using the expression for velocity we can see that $\alpha
=1+\Omega ^{2}$, and the expression for the steady state pulse (\ref{eq13})
coincides with expression (\ref{eq26}).

\section{\protect\LARGE Conclusion}

We have introduced and analyzed a model for the propagation of extremely
short unipolar pulses of an electromagnetic field in a medium represented by
anharmonic oscillators with a cubic nonlinearity. The model under
consideration takes into account the dispersion properties of both linear
and nonlinear responses of the medium. This model is the simplest
generalization of the well known Lorentz model used to describe linear
optical properties in condensed matter. The cubic nonlinearity is the first
type of anharmonic correction to the Lorentz model and it results in the
Duffing oscillator. Here we consider the total Maxwell-Duffing model in
detail. The Lagrangian picture of the TMD model was considered and three
integrals of motion were found. Two families of exact analytical solutions,
with positive and negative polarities, have been found as moving solitary
pulses. The first kind of steady state ESP is an electromagnetic spike
propagating in a nonlinear medium. It was discussed early in \cite
{R2,R7,R8,R14}. A new kind of steady state ESP is an electromagnetic spike
propagating on a non-zero electric background. There are both bright and
dark ESP. Unlike the ESP on a zero background, here pulses of different
polarities have different amplitudes.

We found that the TMD equations can be represented in bilinear Hirota's
form. In the case of a zero background the one-soliton solution of the
bilinear equations was obtained. It coincides with the expression of a
steady state ESP. There are many examples when the bilinear form of the
nonlinear evolution equations leads to the existence of two-soliton
solutions without ensuring complete integrability or/and the existence of $N$%
-soliton solutions. However in this particular case we were not able to
obtain a two-soliton solution of the TMD equations.

\section{\protect\LARGE Acknowledgment}

One of the authors (A.I.M.) is grateful to the \textit{Laboratoire de
Math\'ematiques, INSA de Rouen} for hospitality and support. 


\bigskip

\begin{thebibliography}{99}
\bibitem{R1}  Kazuhiro Akimoto, \textit{J. Phys.Soc.Japan} \textbf{65}, N7,
2020-2032 (1996).

\bibitem{R2}  A. E. Kaplan, S.F. Straub and P. L. Shkolnikov, \textit{%
J.Opt.Soc.Amer.} \textbf{B14}, N11, 3013-3024 (1997).

\bibitem{R3}  N. Bloembergen, \textit{Rev.Mod.Phys}. \textbf{71}, N2,
S283-S287 (1999).

\bibitem{R4}  A.V. Kim, M.Yu. Ryabikin, and A.M. Sergeev, \textit{Uspekhi
Phys. Nauk} \textbf{169}, N1, 58-65 (1999).

\bibitem{R5}  Th. Brabec and F. Krausz, \textit{Rev.Mod.Phys}. \textbf{72},
N2, 545-591 (2000).

\bibitem{R6}  A.D. Vuzha, \textit{Fiz.Tverd.Tela }(Leningrad) \textbf{20},
N1, 272-273 (1978).

\bibitem{R7}  A.I. Maimistov and S.O. Elyutin, \textit{J.Mod.Opt.} \textbf{39%
}, N11, 2201-2208 (1992).

\bibitem{R8}  A. E. Kaplan and P. L. Shkolnikov, \textit{Phys.Rev.Lett}. 
\textbf{75}, N12, 2316-2319 (1995).

\bibitem{Sazo}  S.E. Sazonov, E.V. Trifonov, \textit{J.Phys.} \textbf{B 27},
N1, L7-L12 (1994).

\bibitem{Shvarts}  A.B.Shvartsburg, \textit{Usp.Fis.Nauk} \textbf{168}, N1,
85-103 (1998).

\bibitem{R9}  S.A. Kozlov and S.V. Sazonov, \textit{JETP} \textbf{84}, N2,
221-235 (1997).

\bibitem{R10}  P.A.M. Dirac, \textit{Canad.J.Math}. \textbf{2}, N2, 129-148
(1950).

\bibitem{R11}  C.A.Hurst, \textit{Recent Developm. in Mathemat.Phys}, Eds.
H.Mitter, L.Pittner, Springer-Verlag, Berlin,.(1987), p.18-52.

\bibitem{R12}  R. Hirota, and J. Satsuma, \textit{Progr.Theor.Phys., Suppl}. 
\textbf{59}, 64 (1976).

\bibitem{R13}  M.J. Ablowitz, and H. Segur. \textit{Solitons and the Inverse
Scattering Transform} (SIAM, Philadelphia, 1981).

\bibitem{R14}  A.I. Maimistov, \textit{Quantum Electronics} \textbf{30}, N4,
287-304 (2000).
\end{thebibliography}
\end{document}